\documentclass[12pt]{article}
\usepackage{epsfig}
\renewcommand{\theequation}{\thesection.\arabic{equation}}
\begin{document}

\title{Signature for heavy Majorana neutrinos in hadronic
collisions}
\author{F. M. L. Almeida Jr\thanks{marroqui@if.ufrj.br}, Y. A.
Coutinho\thanks{yara@if.ufrj.br},\\
J. A.  Martins Sim\~oes\thanks{simoes@.if.ufrj.br},
 \\ Instituto de F\'{\i}sica,\\
Universidade Federal do Rio de Janeiro,RJ, Brazil,\\
and\\
M. A. B. do Vale\thanks{aline@if.ufrj.br}\\ 
Centro Brasileiro de Pesquisas F\'{\i}sicas;
\\ Instituto de F\'{\i}sica,\\
Universidade Federal do Rio de Janeiro,\\
RJ, Brazil}

\date{}
\maketitle
\begin{abstract}
\par
 The production and decay of new possible heavy Majorana
 neutrinos are analyzed
 in hadronic collisions. New bounds on the mixing of these
 particles with standard neutrinos are estimated according to a
 fundamental representation suggested by grand unified models.
 A clear signature for these Majorana neutrinos is given by
 same-sign dileptons plus a charged weak vector boson in the
 final state. We discuss the experimental possibilities for the
 future Large Hadron Collider
(LHC) at CERN.
\vskip 1cm
PACS 12.60.-i, 14.60.St
\end{abstract}
\newpage
\section{Introduction}\setcounter{equation}{0}

\par
The increasing experimental evidence for neutrino oscillations
seems to indicate that physics
beyond the standard model may be at sight. The combined results
from solar, atmospheric and laboratory neutrinos, if confirmed,
require at least one new light neutrino \cite{1}. Besides this
possible new fourth neutrino, theoretical models must give some
explanation on the mass spectrum of neutrinos, as well as on the
pattern of mixing angles. This is usually accomplished by
variations of the see-saw mechanism, with new heavy neutral
leptons. Many models that have being proposed recently could
account for these properties but all of them need further
experimental tests \cite{2}. One particular point that still needs
confirmation is the nature of the neutrino fields - Majorana or
Dirac.

\par
In this paper we analyze the production and decay of new heavy
Majorana neutrinos in hadronic collisions. In order to understand
the smallness of neutrino masses, the most likely mechanism
include at least one new heavy Majorana neutrino. These neutrinos
could have its origin in many grand unified models such as in the
fundamental 27-plet of $E_6$ or in left-right symmetric models.
Here we consider the case of new heavy neutral singlets. In
practically all these models we have new interactions, new
fermions and a large Higgs sector. As we have presently no
evidence for new interactions, we will make the hypothesis that
the new neutrino interactions are dominated by the standard Z and
W weak gauge bosons. New Z' and W' are possible but they are
known to have a small coupling with the standard fermions. Since
these Majorana fermions can decay in both charged positive and
negative leptons, they will give us very clear signatures. One
interesting signature is given by the production of a pair of
same-sign dileptons accompanied by a charged weak vector boson.
This signature was already suggested in models with a new right
weak boson \cite {3} and composite Majorana neutrinos \cite {4}.
In a previous work \cite {5} we have presented estimates for this process
in a model with heavy Majorana neutrinos mixed with
standard neutrinos. Here we generalize our results for the mixing
of light and heavy neutrinos and present new bounds on mixing
angles. As we will discuss in section 2, the mixing with new
possible heavy singlets makes pair production of Majorana
neutrinos suppressed relative to a single heavy Majorana
production. We have calculated the general elementary process
$q_i\bar q_j\longrightarrow \ell^-\ell^-W^+$, with $ \ell=e,\mu$,
including finite-width effects, and discuss experimental cuts for
the LHC at CERN energies in section 3 and give our conclusions in
section 4.

\par
\section{The model and bounds on mixing
angles}\setcounter{equation}{0}
\par
The key element in our analysis is the mixing between the
standard light neutrinos and a new heavy one. Recent experimental
searches at LEPII at CERN imply that these new possible neutral
states must have a mass above the standard Z mass \cite {6}. This
kind of mixing is different from the light-to-light mixing, which
seems necessary in order to explain the standard neutrino
properties in solar and atmospheric phenomena. We work in a
scenario where each fermionic family is enlarged and the mixing
occurs with neutrinos of the same generation. In the naive
see-saw model, the mixing between light and heavy neutrinos is
given by $\displaystyle{\theta \simeq {m_{\nu}\over{ m_N}}}$ and
it is clearly out of any visible effect. However, there are many
theoretical models which decouple the mixing from the mass
relation \cite {7}. The mechanism is very simple. In the general
mass matrix including Dirac and Majorana fields one imposes some
internal symmetry that makes the matrix singular. Then the mixing
parameter has an arbitrary value, bounded only by its
phenomenological consequences.
 We take this mixing to be described by a single parameter (for
 each generation), with the new heavy states appearing as
 isosinglets. The general Lagrangian of this kind of model is
 then given by:
\begin{eqnarray}
{\cal L} & = &\sum_{l=e,\mu,\tau}\lbrace-{g\over
2\sqrt2}\Bigl(\bar\nu_l\gamma^{\mu}(1-\gamma^5)l \cos\theta_l+
\bar N_l\gamma^{\mu}(1-\gamma^5)l \sin\theta_l\Bigr)W_{\mu}
\nonumber\\
&-& {g\over
2\cos\theta_W}\Bigl(\bar\nu_l\gamma^{\mu}(g_{V_l}-g_{A_l}
\gamma^5)\nu_l+ {1\over 2}\cos\theta_l
\sin\theta_l \bar N_l\gamma^{\mu}(1-\gamma^5)\nu_l
\nonumber\\
&+&{1\over 2}\sin^2\theta_l \bar N_l\gamma^{\mu}(1-\gamma^5)N_l
\Bigr)Z_{\mu} + h.c.\rbrace ,
\end{eqnarray}
where
\begin{eqnarray}
g_{V_l} & = &g^{SM}_V-{1\over 2}\sin^2\theta_l\nonumber\\
g_{A_l} & = &g^{SM}_A-{1\over 2}\sin^2\theta_l
\end{eqnarray}
\par
We have considered only the first and second families, since
electrons and muons are directly observable particles. The $
\tau^\pm \tau^\pm $ dileptons are also possible, but the
associated Majorana neutrino can be heavier than the
corresponding to the other families. Besides that, the final
state for tau leptonic decays has to be reconstructed with
undetected neutrinos.
\par

 The W and Z couplings to standard leptons are well known to
 agree with the experimental data on low energy lepton-hadron
 scattering, with the high precision Z data at CERN and SLAC, as
 well as with all the charged current experiments. This imply
 that the mixing angle $\theta_\ell$ must be small. It is also
 well known that the present experimental data requires the
 standard model quantum corrections  in order to compare theory
 and data. With this in consideration, we take the changes in the
 physical observables due to new neutrino mixing to be small
 contributions to the standard model theoretical results,
 including first order corrections. With $ \alpha_ {em} $ and
 $M_Z$ as fundamental input parameters, the mixing indicated in
 Eq. (2.1)  above will imply corrections to the effective
 $\mu$-decay constant $G_{\mu}$; charged coupling lepton
 universality, Cabbibo-Kobayashi-Maskawa unitarity and neutral current 
interactions. The most stringent bounds on $\theta_e$ come from the 
effective coupling of the Z to the electron neutrino \cite { 6}
$g_{exp}^{\nu e} =0.528\pm0.085$ and
$\Gamma^{inv}_{exp}(Z)=500.1\pm 1.8$ MeV, to be compared with the
standard model predictions
 $g_{SM}=0.5042$ and $\Gamma^{inv}_{SM}(Z)=501.7\pm0.2$ MeV. For the
 muon neutrino coupling with the Z boson, the Particle Data Group
 quotes $g_{exp}^{\nu \mu} =0.502\pm 0.017$. After a global fit
 to the data we have obtained, at $95\%$ confidence level, the
 upper bounds: 

\begin{eqnarray}
\sin^2 \theta_e < 0.0052 
\end{eqnarray}
\begin{eqnarray}
\sin^2 \theta_\mu < 0.0001 
\end{eqnarray}

\par
These bounds are more restrictive
than earlier analysis \cite {8}. Since the muon bound is much
stronger, the dimuon channel will be much smaller then the
dielectron channel for the LHC at CERN. We are
left then with the possibility of detecting the Majorana
contribution in the $ e^\pm e^\pm $ channel.
\par
With these bounds we can compare the total cross section for
single and heavy pair neutrino production. The interaction shown
in Eq. (2.1) implies that heavy pair production is suppressed if
compared with single production of a heavy neutrino, as it was
shown in \cite {5}

\section{Results}\setcounter{equation}{0}
\par
The fundamental production mechanism is given by quark-antiquark
annihilation into a W followed by the single Majorana production
and decay, as shown in Fig. 1. At the LHC at CERN center of mass
energies ($\sqrt s=14$ TeV) other contributions such as gluon
fusion will give a small contribution. We have also checked the
contribution from W W fusion which turns out to be small. Let us
now discuss the main characteristics of the final state
identification. Experimentaly each charged lepton can be isolated into a cone,
with no accompanying hadrons, in order to eliminate possible
misidentification from hadron decays. The dominant final state
$W\longrightarrow jets$ has a sharp kinematical identification 
in the invariant mass.
The leptonic $W\longrightarrow \ell \nu_{\ell}$ decay can be
identified by the total $p_T$ balance with a reconstructed
neutrino. The proposed detectors \cite {9} for the LHC at CERN
energies are planned to have a good leptonic and hadronic
resolution, as well as a wide rapidity coverage. We have
estimated the total cross section for $pp\longrightarrow
\ell^{\pm}\ell^{\pm}W^{\mp}X$ applying conservative cuts on the
final state. For the charged leptons we considered $p_{T\ell}> 5$
GeV and a pseudo rapidity $|\eta|<2.5$. For the charged weak
boson the cuts $|\eta|<2.5$ and
$p_{TW}> 15$ GeV were done. For the upper bound $\sin^2 \theta_e
= 0.0052$ our results are shown in Fig. 2. The analytic
expression for the elementary process $q_i\bar q_j\longrightarrow
\ell^-\ell^-W^+$ is given in the appendix. For the proton
structure functions we have employed the MRS[G] set \cite{10}.
\par
The standard model background comes mainly from top production and decay 
via the chain $t\longrightarrow b\longrightarrow$ c $\ell \nu_{\ell}$.
It is well known \cite{11} that prompt lepton background from heavy quark 
decay can be suppressed by a suitable isolation criteria on transverse 
energy and momentum. This procedure was applied on earlier analysis 
\cite{12} of Majorana neutrino production.  We employed a cut of $ p_T= 80 $ GeV  on
the softer lepton coming from the standard model background. 
Since our estimate on mixing angles is more restrictive than previous 
values, the signal for Majorana neutrino production will be suppressed. 
An alternative to enhance the signal is to employ a very distinctive 
characteristic between the signal and the standard model background:
whereas in the Majorana neutrino production we have no undetected particles
in the final state, for the heavy quark background we have two undetected
neutrinos in the final state. As a consequence there is no missing transverse 
momentum, $p\!\!\!\slash_T$, 
for the signal and for the background there is an average \cite{12}
$p\!\!\!\slash_T \simeq 50$ GeV. Recent analysis \cite{13} on detector 
possibilities at LHC energies indicate that an error on the missing
transverse energy in the $10-20$ GeV region is expected. The other kinematical 
difference between signal and background is given by the total hadronic invariant 
mass in the final state. In the heavy lepton production we have the associated 
production of a single W boson, with a corresponding peak at the W mass. The 
standard model background has a very different hadronic mass distribution. 
\par
We show in Fig. 3 the convolution for the hadronic invariant mass versus final state missing 
energy variables.  We have applied a cut on the transverse momentum of the least 
energetic lepton of $p_T > 35$ GeV. Both the signal and 
background events were hadronised according to Pythia \cite {14}. The box in Fig. 3 shows 
the kinematical region of the expected signal events. We have considered a heavy
Majorana neutrino with mass of $200$ GeV. From Fig. 2 for a total cross section of 1 fb 
and for the expected luminosity of 100 $fb^{-1}$ at LHC we have a total number of 100 events 
for the signal.
For an invariant hadronic mass in the 60-85 GeV region, a total missing energy less than
12.5 GeV and the most energetic charged lepton transverse momentum greater than 35 GeV,
we still have 62 events from the signal, whereas for the standard model background we 
have only 8 remaining events. With these cuts, we estimate the ratio
$ s/{\sqrt{s+b}}$, where 's' and 'b' are the signal and background number of events, to be 7.4.
 This clearly shows the advantages of applying the above procedure 
for the cuts. For higher neutrino masses the total cross section fall requires more 
stringent cuts on the suggested variables. In this case other kinematical 
restrictions may be useful. The standard model background can still be reduced one order
of magnitude by requiring only two hadronic jets in the final state and by increasing the 
transverse momentum  cut on the charged leptons. In Fig. 4 we display the 
transverse momentum of the final leptons for some values of the Majorana 
neutrino mass. For masses, in the $100-200$ GeV region, a high $p_T$
cut of 80 GeV strongly reduces the signal but for higher masses there are 
practically no changes. For example, for $m_N=200$ GeV the cross section shown in 
Fig. 2 (calculated for $p_T=5$ GeV) is reduced by a factor three if we
increase the $p_T$ cut to 80 GeV. A similar procedure for $m_N=400$ GeV 
reduces the signal by $20 \% $ and for $m_N=800$ GeV there is almost no change.
From the expected luminosity at LHC, we took $m_N=800$ GeV as a superior limit in the 
neutrino mass spectrum. For this extreme value of $m_N$ only one event
is expected, with no further cuts, for a luminosity of 100 $fb^{-1}$. This is a poor 
result from the statistical point of view, but this signal has a very interesting
characteristic. In Fig. 5 we show the distribution for the angle between 
the two final leptons in the laboratory frame. We can see that for heavier 
masses region the two leptons tend to be emitted back-to-back. We can then safely 
expect that the background can be reduced by the simple requirement of low
missing transverse energy in the final state, combined with the hadronic 
invariant mass peaked around the W masss.

\par
We have also estimated the cross section for the Tevatron upgrade
center of mass energies ($\sqrt{s}=2$ TeV) at Fermilab with a
luminosity of 1000 $pb^{-1}$. For a new possible heavy Majorana
neutrino with a mass of 95 GeV we have 10 events applying the following cuts: for 
the leptons and the charged weak bosons we used $|\eta|<2.0$ and $E_l >10$ GeV for 
the leptons, with a mixing parameter $\sin^2 \theta_e=0.0052$. This number falls 
to one for a mass value of 140 GeV doing the same cuts and using the same mixing 
angle.
\par
We now turn our attention to the other kinematical characteristics
of the final state particles. The final W can be identified  in
the channel $ W\longrightarrow jets$ or in the leptonic channel.
The distribution for the total invariant mass $ M_{\ell\ell
W}^2=(p_{\ell}+p_{\ell}+p_W)^2$ is shown in Fig. 6. A better
identification of the heavy neutrino mass is given by the
invariant mass $M_{\ell W}$. Since we have two indistinguishable
particles in the final state, we must take the variable $M_{\ell
W}^2=(p_{\ell}+p_{W})^2$ including all the combinations between
the two leptons and the W. This distribution  presents a well
defined peak at the Majorana mass, as shown in Fig. 7. The total
width for a 100 GeV neutrino is 0.00065 GeV, and for a 800 GeV
neutrino is 0.058 GeV. As mentioned before, another variable that 
can be useful in the
identification of the final state particles is the angle between
the two charged leptons. The distribution for
$\cos\theta_{\ell\ell}$ is shown in Fig. 5, calculated in the
hadronic center of mass frame. For lower $M_N$ mass (100 GeV)
they have a larger probability of being parallel. This is the
case also for the Tevatron at Fermilab with $M_N\approx 90-140$
GeV. For higher masses (800 GeV), they show a preference for
being emitted back-to-back. Finally we show in Fig. 8 the
$p_{TW}$ distribution for the final W. This is an important point
in order to eliminate any possible standard model background. For
lower neutrino mass a cut not greater than 30 GeV must be done in
order not to loose the signal. For higher masses the $p_{TW}$ cut
can be safely higher.

\section{Conclusions}\setcounter{equation}{0}
\par
We have presented an analysis of the possibility of detecting new
heavy Majorana neutrinos through same-sign dileptons and W
detection. The agreement between the standard model predictions
and the present experimental data implies that the mixing angle
between heavy and light states must be small. For the second
family, the upper bound for the mixing angle implies that the
dimuon final state must be suppressed by one order of magnitude relative
to the dielectron channel. For the Tevatron upgrade at
$\sqrt{s}=2$ TeV, we found a detectable signal in a small region in the 
heavy neutrino
mass. For the next LHC the situation is more favorable, allowing
an investigation for new heavy Majorana neutrinos in the range
100-800 GeV. The signal to background ratio can be increased by a
combined cut on the invariant hadronic mass and low missing energy in the final state
instead of a single high $p_T$ cut.
\par

\vskip 4cm

{\it Acknowledgments:} This work was partially supported by the
following Brazilian agencies: CNPq, FUJB, FAPERJ and FINEP.

\vfill\eject
\bigskip

\renewcommand{\theequation}{A.\arabic{equation}}
\LARGE
{\bf Appendix}
\normalsize
\setcounter{equation}{0}
\vskip 1cm
\par
The elementary cross section for the process $q_i
\bar q_j\longrightarrow \ell^-\ell^-W^+$ ($i,j=u,d,s,c$),
can be written as\cite{15}:
\begin{equation}
d\hat \sigma_{ij}=\Delta_{ij}\lbrace D - C \rbrace dLips,
\end{equation}
where
\begin{equation}
\Delta_{ij} ={g^6{U_{ij}}^2  sin^4\theta_{\ell}M_{N\ell}^2\over
6}
\end{equation}
\begin{equation}
D= {2p_{14} p_{23} \over Pr(-p_1-p_2,M_W,\Gamma_W)^2
Pr(p_4+p_5,M_{N\ell},\Gamma_{N\ell})^2}
\end{equation}

\begin{equation}
C={p_{24} p_{13}+ p_{14}p_{23}+
(-p_{14}-p_{23}-p_{24}-p_{13}+p_{12}-{1\over2}M_W^2)p_{12}\over
Pr(-p_1-p_2,M_W,\Gamma_W)^2 Pr(p_4+p_5,M_{N\ell},\Gamma_{N\ell})
Pr(p_3+p_5,M_{N\ell},\Gamma_{N\ell})}
\end{equation}
\bigskip
\par
The propagator effects are included via the function
\begin{equation}
Pr(P, M_k, \Gamma_k)= (P^2-M_k^2) - i(\Gamma_k M_k),\hskip 0.5cm k=W,N_{\ell},
\end{equation}
with $\ell=e,\mu$.
\bigskip
\par
The four-momenta for the quark, antiquark, primary lepton,
secondary lepton and boson are respectively $p_1$, $p_2$, $p_3$,
$p_4$, $p_5$. To simplify our notation we have called the product
between two four-momenta as: $p_{mn}= p_m\cdot p_n$. Finally $U_{ij}$ is the standard 
quark mixing matrix.
\vskip 2cm
\vfill\eject

\vskip 1cm
\vspace{1cm}
\LARGE
{\bf Figure Captions}
\normalsize
\begin{enumerate}
\item Feynman diagram for the elementary process $q_i \bar q_j\longrightarrow 
\ell^-\ell^-W^+$.
\item Total cross section versus $M_{N_e}$ for LHC at CERN and
Tevatron at Fermilab energies.
\item Invariant hadronic mass versus total missing momentum for the Standard Model (crosses) and signal (bullets) events with $M_{N_e}$=200 GeV.
\item Normalized final lepton transverse momentum distributions  
for $M_{N_e}$=100, 200, 400 and 800 GeV for LHC at 14 TeV.
\item Normalized dilepton angular distribution
$\cos\theta_{\ell\ell}$ for $M_{N_e}$=100, 200, 400 and 800 GeV
for LHC at 14 TeV.
\item Normalized invariant mass distribution $M_{\ell\ell W}$ for
$M_{N_e}$=100, 200, 400 and 800 GeV for LHC at 14 TeV.

\item Normalized invariant mass distribution $M_{\ell W}$ for
$M_{N_e}$=100, 200, 400 and 800 GeV
for LHC at 14 TeV.

\item Normalized W transverse momentum distributions for
$M_{N_e}$=100, 200, 400 and 800 GeV
for LHC at 14 TeV.
\end{enumerate}

\begin{figure} [h]
\begin{center}\mbox{\epsfig{file=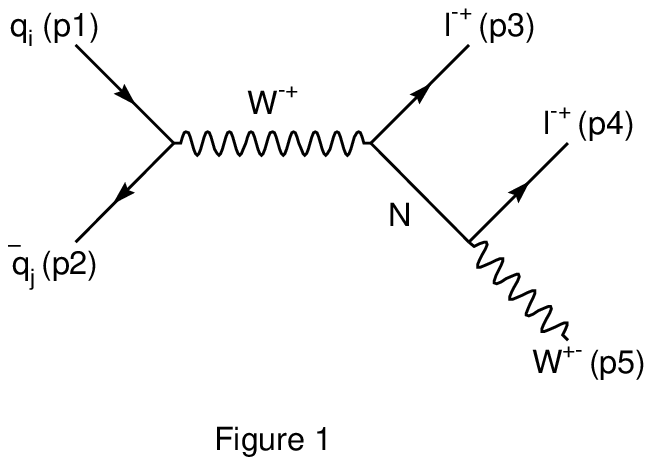,height=6.5cm}}\end{center}
\protect\caption{Feynman diagram for the elementary process $q_i \bar q_j\longrightarrow 
\ell^-\ell^-W^+$.}
\end{figure}

\begin{figure} [h]
\begin{center}\mbox{\epsfig{file=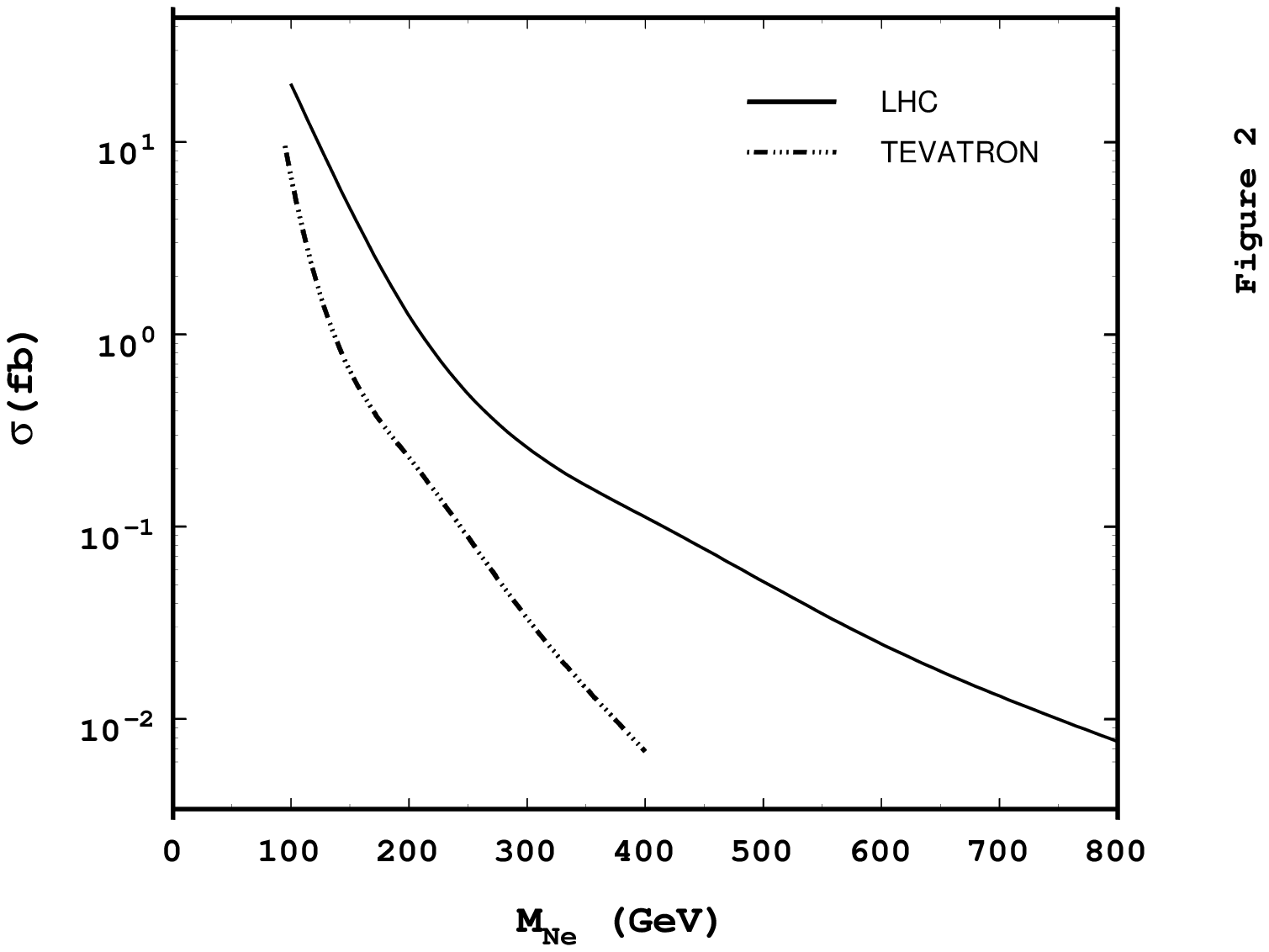,height=6.5cm}}\end{center}
\protect\caption{Total cross section versus $M_{N_e}$ for LHC at CERN and
Tevatron at Fermilab energies.}
\end{figure}

\begin{figure} [h]
\begin{center}\mbox{\epsfig{file=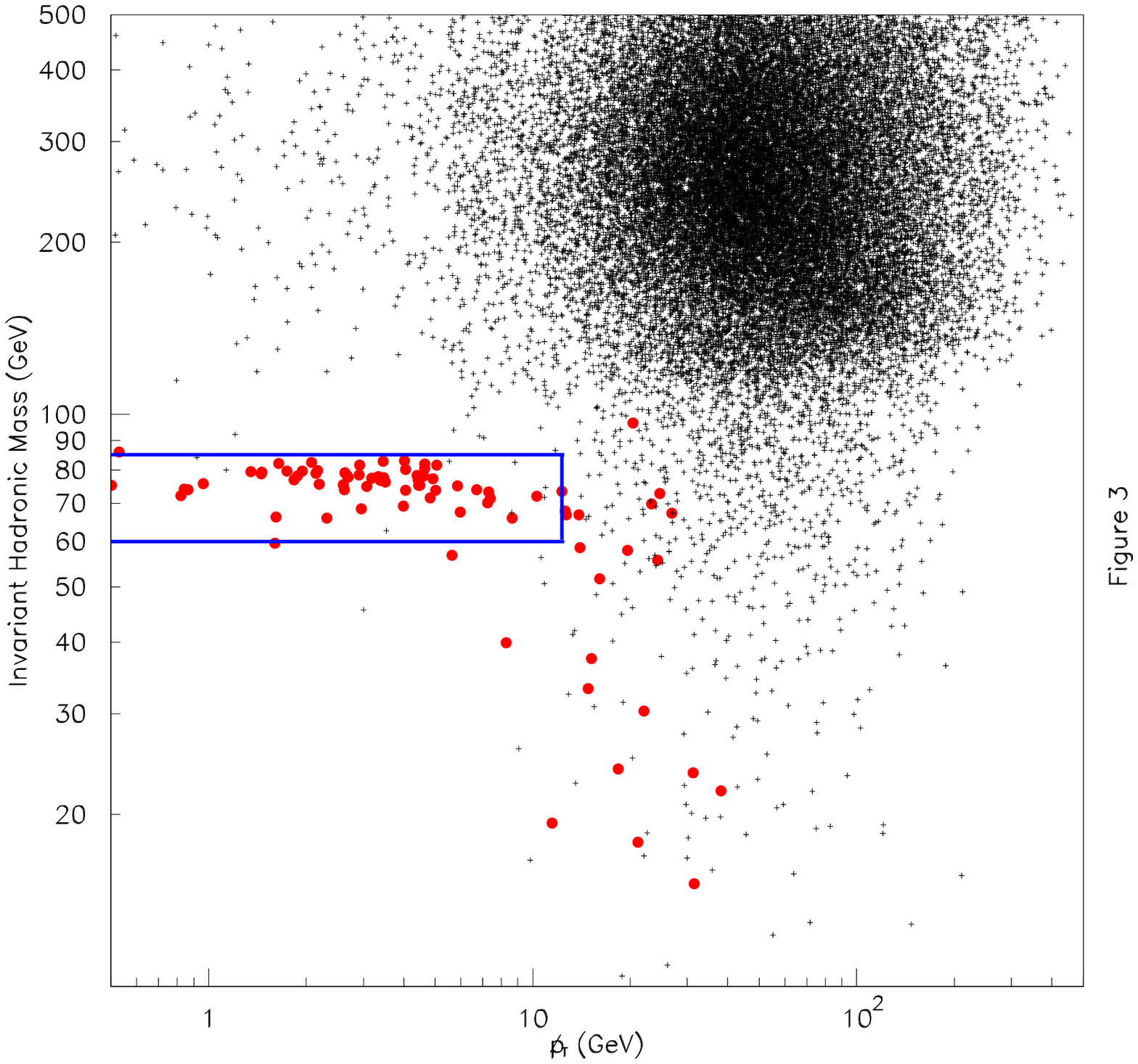,height=6.5cm}}\end{center}
\protect\caption{
Invariant hadronic mass versus total missing momentum for the Standard Model (crosses) and signal (bullets) 
events with $M_{N_e}$=200 GeV.}
\end{figure}

\begin{figure} [h]
\begin{center}\mbox{\epsfig{file=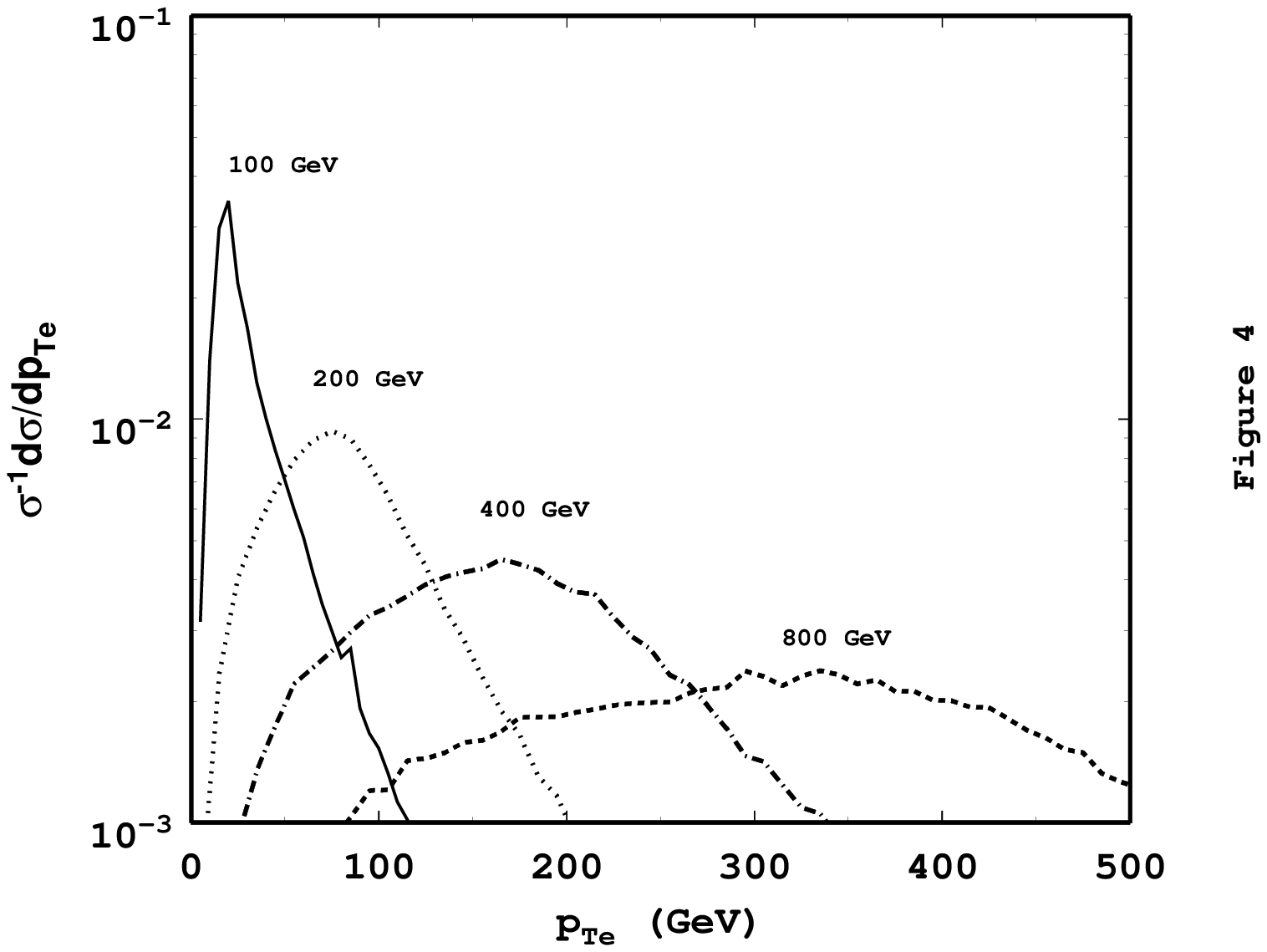,height=6.5cm}}\end{center}
\protect\caption{Normalized final lepton transverse momentum distributions  
for $M_{N_e}$=100, 200, 400 and 800 GeV for LHC at 14 TeV}
\end{figure}

\begin{figure} [h]
\begin{center}\mbox{\epsfig{file=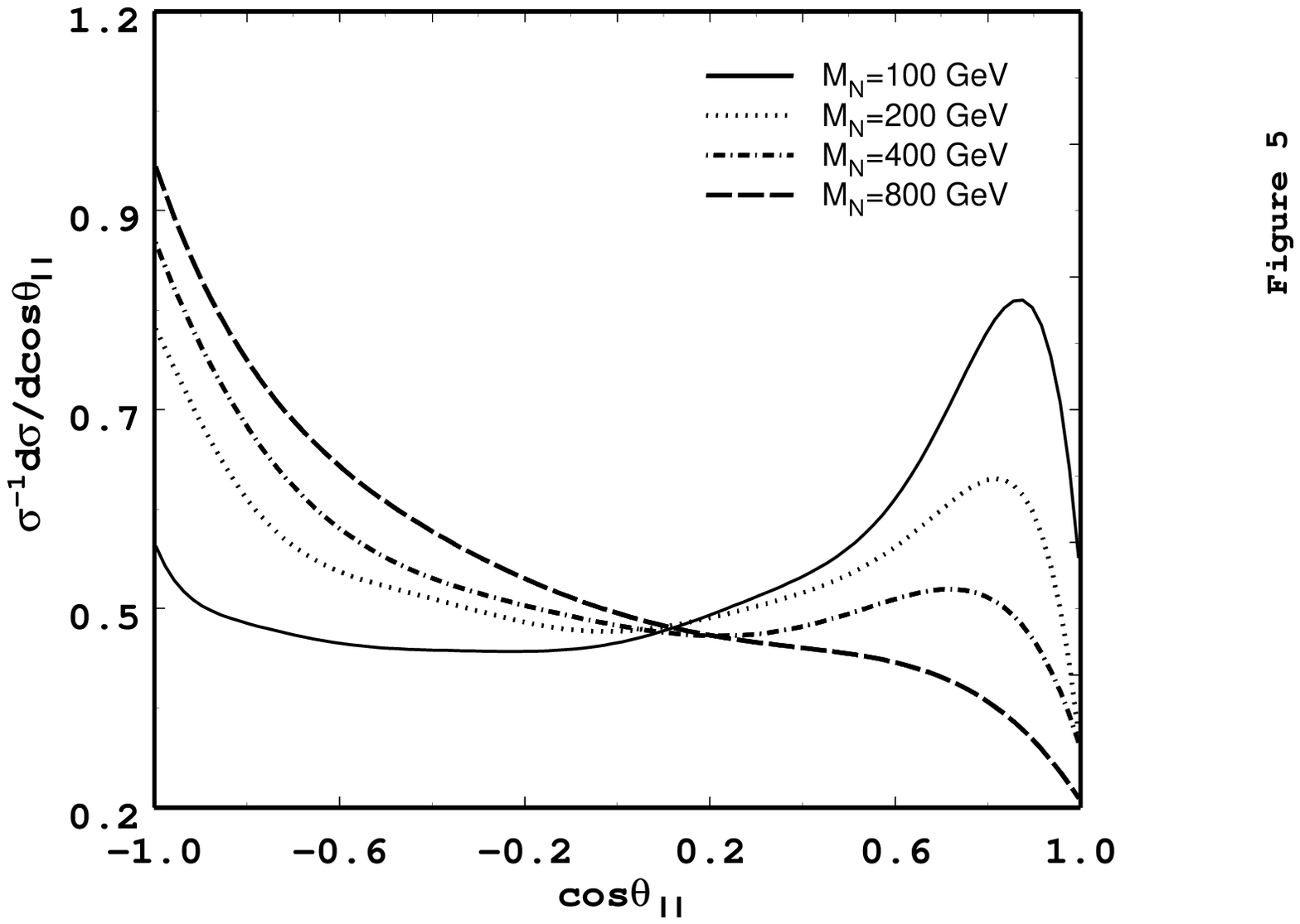,height=6.5cm}}\end{center}
\protect\caption{Normalized dilepton angular distribution
$\cos\theta_{\ell\ell}$ for $M_{N_e}$=100, 200, 400 and 800 GeV
for LHC at 14 TeV.}
\end{figure}

\begin{figure} [h]
\begin{center}\mbox{\epsfig{file=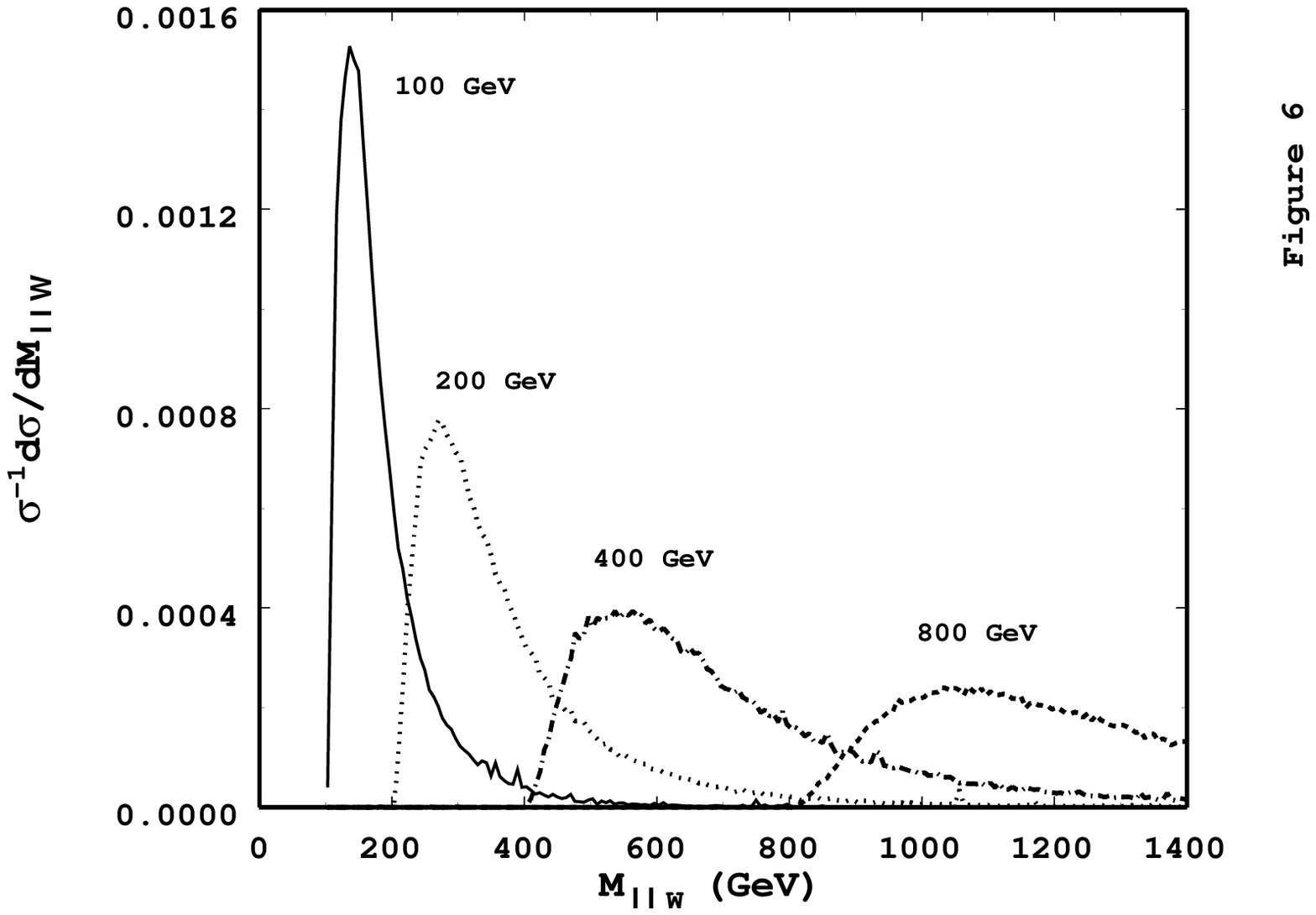,height=6.5cm}}\end{center}
\protect\caption{ Normalized invariant mass distribution $M_{\ell\ell W}$ for
$M_{N_e}$=100, 200, 400 and 800 GeV for LHC at 14 TeV}
\end{figure}

\begin{figure} [h]
\begin{center}\mbox{\epsfig{file=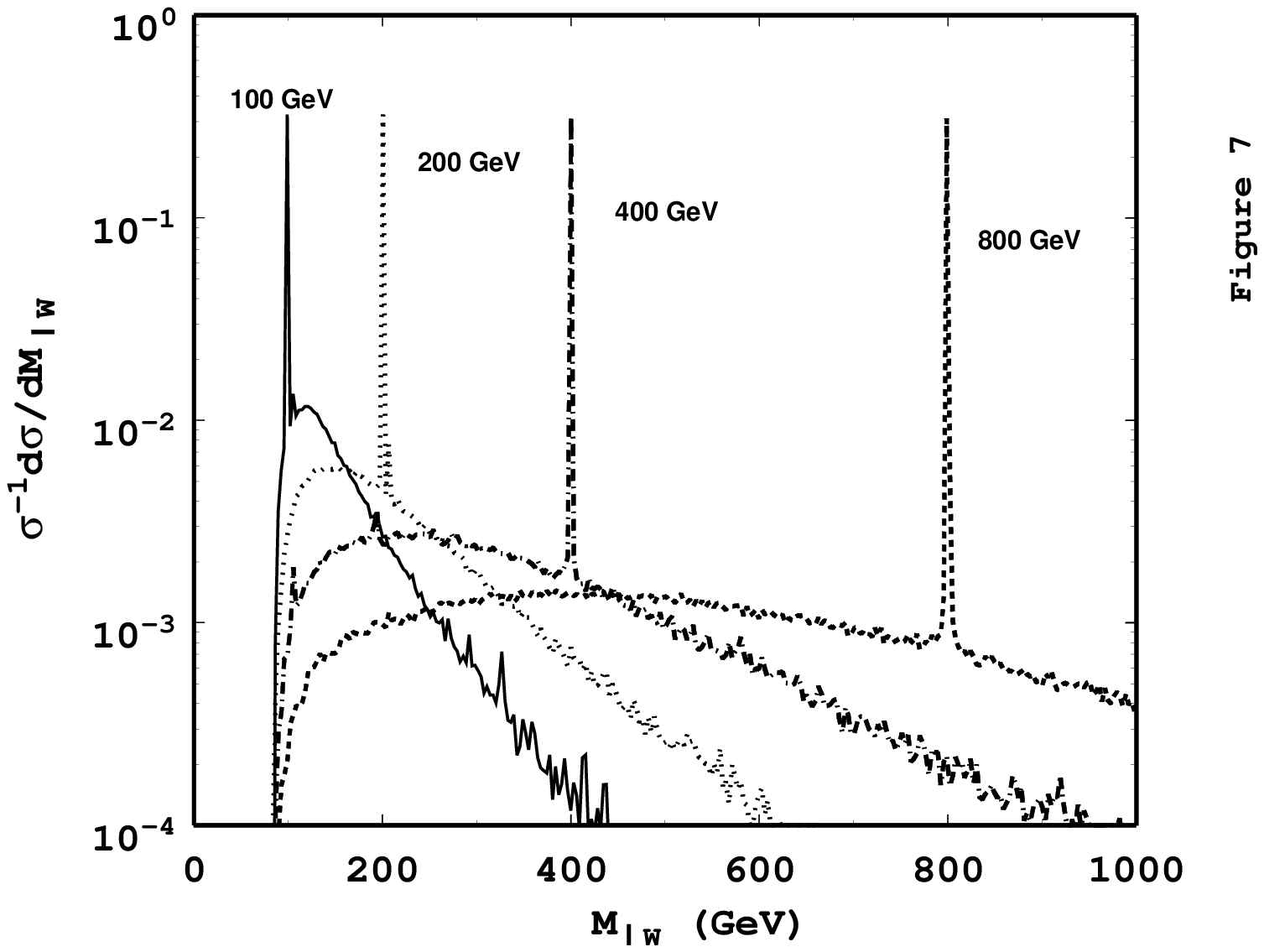,height=6.5cm}}\end{center}
\protect\caption{ Normalized invariant mass distribution $M_{\ell W}$ for
$M_{N_e}$=100, 200, 400 and 800 GeV
for LHC at 14 TeV.}

\end{figure}
\begin{figure} [h]
\begin{center}\mbox{\epsfig{file=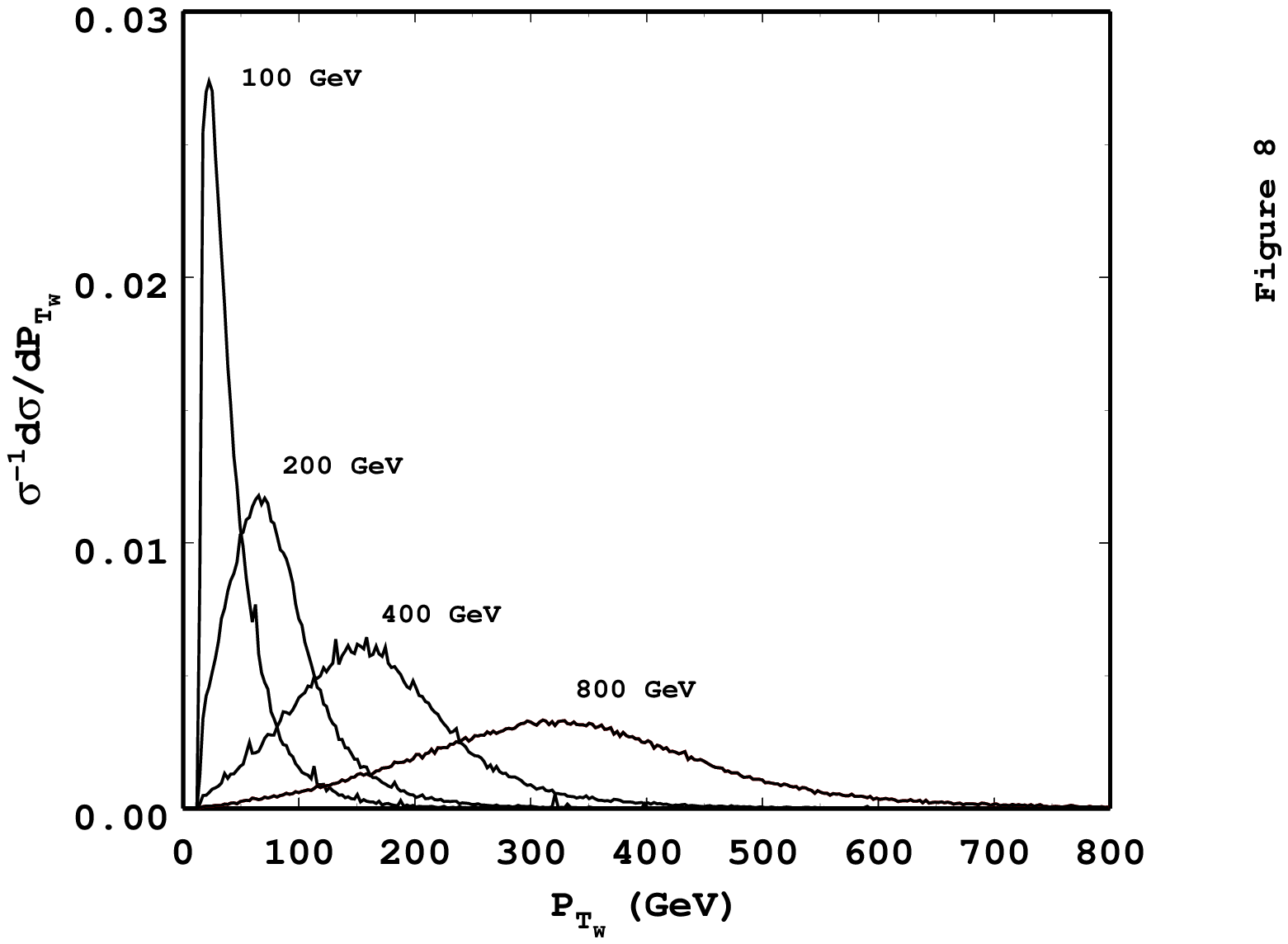,height=6.5cm}}\end{center}
\protect\caption{Normalized W transverse momentum distributions for
$M_{N_e}$=100, 200, 400 and 800 GeV
for LHC at 14 TeV.}
\end{figure}

\end{document}